\documentclass[12pt]{article}

\usepackage{amsmath}
\usepackage{amssymb}

\usepackage{graphicx}

\renewcommand{\thefootnote}{\fnsymbol{footnote}}
\makeatletter \ifcase\@ptsize \font\teneufm=eufm10
\font\seveneufm=eufm7 \font\fiveeufm=eufm5 \font\teneusm=eusm10
\font\seveneusm=eusm7 \font\fiveeusm=eusm5 \or
\font\teneufm=eufm10 scaled \magstephalf \font\seveneufm=eufm7
\font\fiveeufm=eufm5 \font\teneusm=eusm10 scaled \magstephalf
\font\seveneusm=eusm7 \font\fiveeusm=eusm5 \or
\font\teneufm=eufm10 scaled \magstep1 \font\seveneufm=eufm7
\font\fiveeufm=eufm5 \font\teneusm=eusm10 scaled \magstep1
\font\seveneusm=eusm7 \font\fiveeusm=eusm5 \fi

\newfam\eufmfam \newfam\eusmfam \textfont\eufmfam=\teneufm
\scriptfont\eufmfam=\seveneufm \scriptscriptfont\eufmfam=\fiveeufm
\textfont\eusmfam=\teneusm \scriptfont\eusmfam=\seveneusm
\scriptscriptfont\eusmfam=\fiveeusm

\def\frak{\ifmmode\let\next\frak@\else
 \def\next{\errmessage{Use \string\frak\space only in math
 mode}}\fi\next} \def\frak@#1{{\frak@@{#1}}}
 \def\frak@@#1{\fam\eufmfam#1} 
 \def\sh{\ifmmode\let\next\sh@\else
 \def\next{\errmessage{Use \string\sh\space only in math
 mode}}\fi\next} \def\sh@#1{{\sh@@{#1}}}
 \def\sh@@#1{\fam\eusmfam#1}

\ifcase\@ptsize \font\tenmsa=msam10 \font\sevenmsa=msam7
 \font\fivemsa=msam5 \font\tenmsb=msbm10
 \font\sevenmsb=msbm7 \font\fivemsb=msbm5 \or
 \font\tenmsa=msam10 scaled \magstephalf
 \font\sevenmsa=msam7 \font\fivemsa=msam5
 \font\tenmsb=msbm10 scaled \magstephalf
 \font\sevenmsb=msbm7 \font\fivemsb=msbm5 \or
 \font\tenmsa=msam10 scaled \magstep1 \font\sevenmsa=msam7
 \font\fivemsa=msam5 \font\tenmsb=msbm10 scaled \magstep1
 \font\sevenmsb=msbm7 \font\fivemsb=msbm5 \fi

\newfam\msafam \newfam\msbfam \textfont\msafam=\tenmsa
\scriptfont\msafam=\sevenmsa \scriptscriptfont\msafam=\fivemsa
\textfont\msbfam=\tenmsb \scriptfont\msbfam=\sevenmsb
\scriptscriptfont\msbfam=\fivemsb

\def\Bbb{\ifmmode\let\next\Bbb@\else
 \def\next{\errmessage{Use \string\Bbb\space only in math
 mode}}\fi\next} \def\Bbb@#1{{\Bbb@@{#1}}}
 \def\Bbb@@#1{\fam\msbfam#1} \def\hexnumber@#1{\ifnum#1<10
 \number#1\else \ifnum#1=10 A\else\ifnum#1=11
 B\else\ifnum#1=12 C\else \ifnum#1=13 D\else\ifnum#1=14
 E\else\ifnum#1=15 F\fi\fi\fi\fi\fi\fi\fi}
 \def\msa@{\hexnumber@\msafam} \def\msb@{\hexnumber@\msbfam}
 \mathchardef\square="0\msa@03

\makeatother
\newcommand{\beq}{\begin{equation}}
\newcommand{\eeq}{\end{equation}}
\newcommand{\ba}{\begin{array}}
\newcommand{\ea}{\end{array}}
\newcommand{\bea}{\begin{eqnarray}}
\newcommand{\eea}{\end{eqnarray}}
\newcommand{\bean}{\begin{eqnarray*}}
\newcommand{\eean}{\end{eqnarray*}}
\def\beqa{\begin{eqnarray}}
\def\eeqa{\end{eqnarray}}

\newcommand{\be}{\begin{equation}}
\newcommand{\ee}{\end{equation}}
\newtheorem{theorem}{Theorem}[section]

\newtheorem{remark}[theorem]{Remark}

\newtheorem{proof}{Proof.}

\makeatother
 \newcommand{\RR}{{\Bbb R}}
\newcommand{\CC}{{\Bbb C}} 
\newcommand{\ZZ}{{\Bbb Z}} \newcommand{\QQ}{{\Bbb Q}}
 \newcommand{\NN}{{\Bbb N}}

 \def\be{\beta}

\begin{document}

\begin{titlepage}

%\hfill{DFPD01/TH/18}

\hfill{hep-th/}

\vspace{1cm}
\begin{center}
{\Large \bf Zamolodchikov operator-valued relations for \\ $SL(2,R)_k$ WZNW model \\}

%\vspace{.3cm}
%{\large \bf over Special Riemann Surfaces}

\end{center}
\vspace{1.5cm} \centerline{Gaetano Bertoldi \,and\, Gast\'on Giribet} \vspace{0.8cm}
\centerline{\it Institute for Advanced Study} 
\centerline{\it Einstein Drive, Princeton NJ 08540} 

\vspace{2cm} \centerline{\sc ABSTRACT}

\vspace{0.6cm} \noindent

An infinite set of operator-valued relations that hold for reducible representations 
of the $\hat {sl(2)}_k$ algebra is derived. These relations are analogous to those recently
obtained by Zamolodchikov which involve logarithmic fields associated to the Virasoro degenerate representations in Liouville theory. The fusion rules of the $\hat {sl(2)} _k$ 
algebra turn out to be a crucial step in the analysis. The possible relevance of these relations for the boundary theory in the $AdS_3/CFT_2$ correspondence is suggested.

\vspace{0.6cm} \noindent

\end{titlepage}

\newpage

\setcounter{footnote}{0}

\renewcommand{\thefootnote}{\arabic{footnote}}

\renewcommand{\theequation}{\thesection.\arabic{equation}}
\newcommand{\mysection}[1]{\setcounter{equation}{0}\section{#1}}

%%%%%%%%%%%%%%%%%%%%%%%%
\section{Introduction}

In \cite{Zamo}, Al. Zamolodchikov proved the existence
of a set of operator-valued relations in Liouville field theory.
There is one such relation for every degenerate Virasoro primary field, 
which is labelled by a pair of positive integers $(m,n)$. 
These relations correspond to a higher order generalization of the Liouville equation of motion.

In this letter, we prove that a similar set of relations holds for Kac-Moody degenerate representations 
in conformal theories with $\hat {sl(2)} _k$ affine symmetry.  
These operator-valued identities are translated into differential equations satisfied 
by correlators involving states of reducible representations in the $SL(2,\RR )_k$ WZNW model. 
These are differential equations in terms of the ($x,\bar x $) $SL(2,\RR )$-isospin variables.
%%%%%%%%%%%%%%%%%%
%\noindent
Schematically
\medskip
\begin{equation}
\bar K_{m,n} K_{m,n} 
\Phi'_{j_{m,n}}(x,\bar x| z, \bar z) 
= B_{m,n} \Phi_{\tilde \jmath_{m,n}}(x,\bar x| z, \bar z)\,, 
%\label{higher1}
\end{equation}
%\medskip
where $\Phi'_{j_{m,n}}$ are the logarithmic fields associated to
the degenerate Kac-Moody primaries $\Phi_{j_{m,n}}$.
By $\bar K_{m,n}$ and $K_{m,n}$ we denote the left and right operators
that create the null vector in the Verma module generated 
by $\Phi_{j_{m,n}}$. Finally, $\Phi_{\tilde \jmath_{m,n}}$ is
a Kac-Moody primary and $B_{m,n}$ is a $c$-number we will call
{\it Zamolodchikov coefficient}.

The WZNW model on $SL(2,\RR )$ represents the worldsheet CFT describing string theory on $AdS_3$ 
target space. This model has been extensively studied in this context and in relation 
to exact string backgrounds. The string spectrum in $AdS_3$ is constructed in terms of continuous 
and discrete irreducible representations of $SL(2,\RR )_k$ (see \cite{malda} and references therein); 
these representations are typically classified by a complex number
\footnote{See below for detailed description of the representations.} $j$.  
In particular, unitarity of the theory requires a truncation of the set of 
discrete representations which is done by imposing the bound
\begin{equation}
\frac {1-k}{2} < j < -\frac 12\,.  \label{bound}
\end{equation}
The consideration of three and four-point functions leads to even more 
restrictive constraints on the set of states involved in the correlators.

Here, we focus our attention on reducible representations of $SL(2,\RR )_k$. 
These representations are classified by special values of the index 
$j=j^{\pm}_{m,n}$, namely
\begin{equation}
j^+ _{m,n} = \frac {m-1}{2} + \frac {n-1}{2} (k-2) \ , 
\  \ \ j^- _{m,n} = -\frac {m+1}{2} - \frac {n}{2} (k-2)\,,  \label{u}
\end{equation}
for $(m,n)$ a pair of positive integers.

Certainly, these values of $j$
%$j^{\pm}_{m,n}$ 
do not belong to the range (\ref{bound}) and therefore do not represent
perturbative string states in $AdS_3$. However, states belonging to (\ref{u}) 
were previously considered in this theory; 
%within the context of this theory; 
for instance, four-point functions involving the state $j^- _{1,1}=-\frac k2$ 
were considered in \cite{malda} in order to construct three-point string amplitudes 
violating the winding number. The corresponding vertex operator $\Phi _{j^- _{1,1}}$ 
was referred to as the {\it spectral flow operator}. 
Also in \cite{Gaston1} the highest-weight state of this representation 
was considered as the conjugate representation of the identity operator $\Phi _{j^+_{1,1}}$. 

In \cite{Andreev}, the admissible representations of $SL(2,\RR )_k$ were studied 
in relation to a certain identity existing between correlators of WZNW theories 
and minimal models. By analyzing the correlators, 
Andreev was able to reobtain the fusion rules for these representations 
originally found by Awata and Yamada in \cite{AwataYamada}.

The states $\Phi _{j^+_{3,2}}$, $\Phi _{j^+_{2,2}}$ and $\Phi _{j^+_{2,1}}$ 
were considered in \cite{ponsot} as particular examples to discuss the explicit form 
of the four-point functions in $\hat {sl(2)}_k$ and $\hat {su(2)}_k$ models. 
Four-point functions involving generic states of admissible representations 
were also discussed in detail in references \cite{petko}. 
Furthermore, the representation $j^+ _{2,1}$ was employed by Teschner to compute three-point 
functions in $SL(2,\CC )/SU(2)$ WZNW model by using the boostrap approach \cite{teschner}; 
the representation $j^+ _{1,2}$ was also discussed there. The field $\Phi _{j^+_{2,1}}$ was also studied in \cite{satoh} in the context of the path integral approach to string theory in $AdS_3$. 

These representations were also considered in reference \cite{branas} in the context 
of $D$-branes in $AdS_3$ string theory, where solutions analogous to the ZZ branes of 
Liouville theory were discussed (see \cite{branas2} for related discussions).

%***************

Notice that only the values $j^+ _{m,1}$ are $k$-independent
%\footnote{On the other hand, certain $k$-dependent conditions for values of $j$ 
%turn out to be relevant for the description of instantonic configurations 
%in the string worldsheet description.}
and therefore finite in the classical limit, $k \rightarrow \infty$.
We will refer to these particular representations  
as the {\it classical branch}. 
In this particular case, %where a sensible classical limit exists,
it is possible to verify directly that the 
above relations hold at the classical level.
%the operator-valued relations in terms of functional properties of wave functions on $SL(2,\RR )$.

The paper is organized as follows: In section 2, we discuss the $\hat{sl(2)} _k$ affine algebra 
and its unitary representations. We review the formula for three-point correlation functions 
and we list some facts about reducible representations. 
In section 3, we introduce logarithmic operators in the CFT and provide a
preliminary argument that the above operator-valued relations hold. 
We also show that they are verified at the classical level for 
the operators in the {\it classical branch}. 
Section 4 contains a complete proof and the   
explicit formulas for the Zamolodchikov coefficients $B^\pm_{m,n}$.
We dedicate section 5 to the discussion of the results.

\section{The $\hat {sl(2)} _k$ affine algebra}

\subsection{Kac-Moody algebra and unitary representations}

The $\hat {sl(2)}_k$ affine algebra is defined by the following Lie products
\begin{eqnarray}
[ J^3_n ,J^{\pm}_m ]= \pm J^{\pm}_{n+m} , \ \ [ J^3_n ,J^3_m ] = - \frac k2 n \delta _{n,-m} , \ \ [ J^+_n ,J^-_m ]= 2 J^3_{n+m} - kn\delta _{n,-m}  \label{unpebete}
\end{eqnarray}
By defining the operators
\begin{eqnarray*}
L_n = - \frac {1}{k-2} \sum _{n\in \ZZ} :J^+_n J^- _{m-n}+J^-_n J^+ _{m-n}+2J^3_nJ^3_{m-n}:  
\end{eqnarray*}
it is feasible to show that we obtain a representation of the 
Virasoro algebra with central charge $c$ given by
\[
c=3 + \frac {6}{k-2}
\]
Here, we will consider the case $k>2$.

As usual, we can encode the $\hat {sl(2)} _k$ structure in the operator product 
expansion of local operators $J^a (z) = \sum _{n\in {\ZZ}} z^{-n-1} J^a_n$, $a = \{+,-,3\}$.

%\subsection{Unitary representations}
Representations of $SL(2,\RR)$ are classified 
by an index $j$ and the vectors $| j, m \rangle$ 
are labelled by an index $m$. 
Hermitian unitary representations are listed below:

{\it Highest weight discrete series ${\cal D} ^{+}_j$:} 
In these infinite dimensional representations $2 j\in {\NN}$ 
and $m\in j-\NN $.

{\it Lowest weight discrete series ${\cal D} ^{-}_j$:} 
These are analogous to the highest weigth series, being $2j\in {\NN}$ 
and $m\in \NN-j$.

{\it Complementary series ${\cal E} ^{\alpha }_j$:} 
These are defined for $j \in (\frac 12 ,1)$ and 
$2 j> - 1 - \left| 2 \alpha - 1 \right|$, where $\alpha \in (0,1]$ and $m\in \alpha + \ZZ$.

{\it Principal continuous series ${\cal C} ^{\alpha }_{t}$:} 
These are defined for $j\in -\frac 12 +i t$, with 
$t\in \RR$, $\alpha \in (0,1]$ and  $m\in \alpha + \ZZ$.

{\it Identity representation ${\cal I}$:} This is defined for $j=m=0$.

The spectrum of $j$ can be restricted to $j<-\frac 12$ by considering the 
invariance under Weyl reflection $j \leftrightarrow -1-j$. Moreover, 
the indices of discrete representations ${\cal D} ^{\pm} _j$ 
have to be bounded from below as $\frac {1-k}{2}<j$ in order to guarantee 
the non-negative norm condition for the states of the Kac-Moody module. 

The representations of the universal covering of $SL(2,\RR)$ are defined by the 
relaxation of the condition $2j\in \NN \rightarrow \RR$ 
for the discrete series ${\cal D}^{\pm }_{j}$.

The Kac-Moody primary states $\left| j,m \right>$ are classified 
in terms of the mentioned representations and satisfy the following properties
\begin{eqnarray}
J^3 _0 \left| j,m \right> = m \left| j,m \right> \ , \ \ \ J^{\pm} _0 \left| j,m \right> = 
(\mp j-m) \left| j,m \pm 1 \right> \ , \ \ \ J^{a}_{n>0} \left| j,m \right> = 0  \label{kmprimarios}
\end{eqnarray}
where $a = \{+,-,3\}$.
These states are primary states of the Virasoro algebra, namely
\begin{eqnarray*}
L_0 \left| j,m \right> = \Delta _j \left| j,m \right> \ , \ \ \  L_{n>0} \left| j,m \right> = 0
\end{eqnarray*}
where the $L_n$ are the Fourier modes of the Sugawara stress-tensor 
$T(z) = \sum _{n\in {\ZZ}} z^{-n-2} L_n$ and $\Delta _j = -\frac {j(j+1)}{k-2}$. 
Notice that the spectrum (which basically is given by the value of the quadratic Casimir) 
remains invariant under Weyl reflection $j \rightarrow \sigma ^+ (j)=-1-j$. 
Besides, the quantity $\Delta _j -j$ is invariant under 
the {\it second} Weyl reflection $j \rightarrow {\sigma } ^- (j) =1-k-j$.

%\paragraph{Spectral flow symmetry.}
The $\hat {sl(2)}_k$ algebra (\ref{unpebete}) remains invariant under the following transformations
\begin{eqnarray*}
J^{\pm} _n \rightarrow \tilde {J} ^{\pm} _n = {J} ^{\pm} _{n\pm \omega} \ , \ \ \ J^3 _0 
\rightarrow \tilde {J} ^3 _0 ={J} ^3 _0 + \omega \frac k2
\end{eqnarray*}
for $\omega \in \ZZ$. This automorphism is called {\it spectral flow} and generates 
an infinite set of new representations, which we denote as ${\cal D} ^{\pm, \omega } _j, 
{\cal C} ^{\alpha , \omega }_{t}$ and $ {\cal E} ^{\alpha ,\omega }_{j} $ 
with the intention to explicitly refer to the sector $\omega $ containing them 
as proper primary states. More precisely, the states of these {\it new} 
representations can be labelled by an additional index $\omega$ and are given 
by vectors which are Kac-Moody primaries 
with respect to the {\it new} generators ({\it i.e.} the $\tilde {J} ^a _n$). 
These states are also primary states in the original Virasoro algebra, satisfying
\begin{eqnarray}
L_0 \left| j,m,\omega \right> = \Delta _j ^{\omega} \left| j,m,\omega \right> \ , 
\ \ \  L_{n>0} \left| j,m \right> = 0
\end{eqnarray}
where $\Delta _j ^{\omega } = \Delta _j -m\omega -\frac k4 \omega ^2$.
Then, the Hilbert space is parametrized by a set of three quantum numbers, 
namely $\left| j,m,\omega \right>$.

It is worth remarking that certain states belonging to the representations 
${\cal D} ^{\pm, \omega = 0 } _j$ are identified with states of the representation 
${\cal D} ^{\mp, \omega = \pm 1 } _{-\frac k2 -j}$, since the states of the 
form $\left| j,\mp j,0 \right>$ coincide with $\left| -\frac k2 -j,\mp \frac k2\mp j,\pm 1  \right>$. 
Let us finally introduce $\sigma ^0 (j) = -\frac k2 -j$, satisfying 
\begin{equation}
\frac 12 [\sigma ^- ,\sigma ^+ ] (j)  = \pm [\sigma ^{\pm} , \sigma ^0 ] (j) =  k-2\,.
\end{equation}
Observe that the three transformations $\sigma ^a$ coincide 
in the {\it tensionless} limit $k \rightarrow 2$.

\subsection{Vertex operator algebra}

The vertex operators $\Phi _{j,m} (z)$ create the states $\left| j,m \right>$ from 
the $SL(2,\RR)$ invariant vacuum $\left| 0 \right>$; these are defined by the action
\begin{eqnarray}
\lim _{z \rightarrow 0} \Phi _{j,m} (z) \left| 0 \right> =  \left| j,m \right> 
\end{eqnarray}
These are local operators which can be associated to differentiable 
functions $\Phi _{j} (z|x)$ on the manifold. 
We have that
\begin{equation}
[ J^a _n,\Phi _{j} (z|x) ] = z^n D^a _j \Phi _{j} (z|x)
\label{hey1}
\end{equation}
 where $a = \{+,-,3\}$ and 
\begin{equation}
D^3 = x\partial _x -j, \ \ \ D^- = -\partial _x, \ \ \ D^+ = -x^2 \partial _x +2jx
\label{hey2}
\end{equation}
form a representation of $sl(2,\RR)$. 

Then, the eigenfunctions $\psi ^a _{j,m} (x)$ of the operators $ D ^a$ are given by
\[
\psi ^3 _{j,m} (x ) = x ^{j+m} \ , \ \ \ \psi ^{\pm} _{j,m} (x ) = x ^{j \mp j } e^{\pm m x ^{\pm 1}}\,.
\]
Let us consider $\psi ^3 _{j,m} (x)$. It is easy to show that this basis 
corresponds to the $\hat {sl(2)} _k$ block structure for the states 
$\left| j,m \right>$ presented in (\ref{kmprimarios}). 
In particular, the zero modes satisfy the following product
\begin{eqnarray*}
[J^3_0,\Phi _{j,m} ] = m \Phi _{j,m} , \ \ \ [J^{\pm}_0,\Phi _{j,m} ] = (\mp j - m) \Phi _{j,m\pm 1} 
\end{eqnarray*}
which mimics the fact that 
\begin{equation}
D^3 \psi ^3 _{j,m} = m \psi ^3 _{j,m} \ , \ \ \ 
D^{\pm} \psi ^3 _{j,m} = (\mp j - m) \psi ^3 _{j,m\pm 1} \ , \ \ \ 
D^{\pm } \psi ^{\pm} _{j,m} = m \psi ^{\pm} _{j,m} \,.
\end{equation}
Then, the representations $\Phi _{j} (z|x)$ are given by meromorphic functions in $x$ 
and can be considered as the Fourier transform of the representations $\Phi _{j,m} (z)$, 
which, by using the basis of eigenfunctions $\psi ^3_{j,m} (x)$, 
can be written as the following spectral decomposition
\begin{equation}
\Phi _{j,m,\bar m} (z,\bar z) = \int d^2 x\, x^{j+m} \bar x ^{j+\bar m} \Phi _{j} (x,\bar x | z,\bar z)
\end{equation}
%*******
where we have explicitly considered the antiholomorphic part of the 
$\hat {sl(2)} _k \otimes \bar {\hat {sl(2)}} _k$ algebra. 
%consequently and, within the context of applications to string theory, 
In the context of the $AdS_3 / CFT_2$ correspondence, 
$\Phi _{j} (z|x)$ corresponds to a bulk-boundary propagator \cite{ooguri}, 
%in the $AdS_3 /CFT_2$ correspondence, 
where ($x,\bar x$) are the coordinates of the space 
where the dual BCFT is formulated.

%*******

The $\Phi _j (x|z)$ are associated to differentiable functions on the group manifold.
In the study of the $SL(2,\RR )_k$ WZNW model, 
since one has no direct access to them as in the case of models 
corresponding to euclidean target spaces, 
it is usual to investigate the properties of the observables by 
considering the analytic continuation of the model on $SL(2,\CC )/SU(2)$. 
An example of this is the study of the two and three-point functions in string theory 
on $AdS_3$ \cite{malda}, where the states of the model on $SL(2,\RR )$ appear as 
pole conditions of the analytic extension of the results 
obtained for the euclidean model $SL(2,\CC )/SU(2)$. 
Likewise, vertex operators are constructed by analytic 
continuation of the wave functions in the homogeneous space $SL(2,\CC )/SU(2)$.

A convenient representation for these wave functions can be given 
in terms of the Gauss parametrization of the group elements, namely
\begin{equation}
\Phi _j (x |z) = \frac {2j+1}{\pi } \left( |\gamma -x|^2 e^{\phi } + e^{-\phi } \right) ^{2j}
\label{cipollina}
\end{equation} 
where $\gamma \in \CC$ and $\phi \in \RR $. In the quantum case, 
$\phi $ receives corrections as $\phi \rightarrow \phi / \sqrt {k-2}$.

Next, we will introduce the reducible representations of
$SL(2,\RR )_k$, which are the central element of the discussion. 
%%%%%%%%%%%%%%%%%%%%%%%%%%%%%%%%%%%%%%%%%%%%%%%%%%%%

%%%%%%%%%%%%%%%%%%%%%%%%%%%%%%%%%%%%%%%
\subsection{Degenerate representations}

Kac and Kazhdan \cite{KacKazhdan} found that a highest weight representation
of $sl(2)_k$ is reducible if the highest weight $j$ takes the values 
\begin{equation}
j^+_{m,n} = \frac{m-1}{2} + \frac{n-1}{2}(k-2)\,, \quad
j^-_{m,n} = - \frac{m+1}{2} -\frac{n}{2}(k-2)\,. \label{arr}\,
\end{equation}
In particular, there exists a null vector
$| \chi^{\pm}_{m,n} \rangle$ with dimension 
$$
\Delta^+_{m,n} = \Delta_{m,n} + m (n-1)\,,
\quad
\Delta^-_{m,n} = \Delta_{m,n} + m n \,, 
$$
and charge
\begin{equation}
\tilde \jmath^+_{m,n} = j^+_{m,n} - m
=  -\frac{m+1}{2} + \frac{n-1}{2}(k-2)
\,,
\quad
\tilde \jmath^{\,-}_{m,n} = j^-_{m,n} + m
=
\frac{m-1}{2} -\frac{n}{2}(k-2)\, 
\,,
\label{tildejmn}\end{equation}
respectively. 
%(see \cite{seiberg} for interesting discussion about Liouville degenerate representations).
More precisely, the null states that are present in the Verma module
\footnote{degenerate representations.} of these representations are given by 
\cite{KacKazhdan,MalikovFF}
\begin{equation}
\left| \chi _{n,m} ^{\pm }\right>= \bar {K} ^{\pm} _{m,n} K^{\pm} _{m,n} \left| j^{\pm} _{m,n} \right>  \label{moco}
\end{equation}
where the {\it decoupling} operators $K^{\pm }_{m,n}$ can be written as follows
\begin{eqnarray}
K^{+} _{m,n} = \left( J ^- _0 \right) ^{m-(n-1)(k-2)} \left( J ^+ _{-1} \right) ^{m-(n-2)(k-2)} ...  
\left( J ^+ _{-1} \right) ^{m+(n-2)(k-2)} \left( J ^- _0 \right) ^{m+(n-1)(k-2)}  \\
K^{-} _{m,n} =  \left( J ^+ _{-1} \right) ^{m-(n-1)(k-2)} \left( J ^- _{0} \right) ^{m-(n-2)(k-2)} 
...  \left( J ^- _{0} \right) ^{m+(n-2)(k-2)} \left( J ^+ _{-1} \right) ^{m+(n-1)(k-2)} \label{so}
\end{eqnarray}
By (\ref{moco})-(\ref{so}) and (\ref{hey1},\ref{hey2}), the decoupling conditions for null states 
translate into differential equations to be satisfied by correlation functions involving 
$\Phi _{j^{\pm }_{m,n}}$.

As we have already mentioned in the introduction, we will refer to the 
states labelled by  
$j^+_{m,1}$ as the {\it classical branch}, because they
have a classical limit. In this branch, the {\it decoupling} differential 
equations simply reflect the fact that the wave functions corresponding
to $j^{\pm }_{m,1}$ are polynomials in the ($x,\bar x $) coordinates.
In fact 
\footnote{where the l.h.s. of (\ref{tuluca}) has to be understood 
schematically as representing decoupling conditions of null states.}
\begin{equation}
\bar {K} ^{+} _{m,1} K^{+} _{m,1} \left| j^+_{m,1} \right>= 0 \ \ \  \rightarrow 
\ \ \ \  \partial ^m _{\bar x } \partial ^m _x \Phi _{\frac {m-1}{2}} (x|z)  
= 0 \,. 
\label{tuluca}
\end{equation}
\noindent
Let us also notice the following properties holding for degenerate representations
\begin{eqnarray}
\tilde \jmath ^{\pm} _{m,1} = \sigma ^{\pm } (j^{\pm} _{m,1})  \ , \ \ \ \  
j ^{\pm } _{m,n} = \sigma ^0  (j^{\mp } _{m,n} ) \,.
\end{eqnarray}
Reducible representations are the fundamental elements in our discussion. 
The other ingredient which turns out to be important in the analysis is the 
expression of three-point correlation function,
which is given in the following subsection.

%%%%%%%%%%%%%%%%%%%%%%%%%%%%%%%%%%
\subsection{Three-point correlation functions}

The expressions of two and three-point functions in the gauged $SL(2,\CC )/SU(2)$ 
WZNW model were computed in \cite{teschner}. In \cite{malda}, the interpretation 
of these correlators as those describing string scattering amplitudes 
in $AdS_3$ was carefully carried out. For a complete discussion 
of three-point functions in $SL(2,\RR )$ see \cite{Gaston1,malda}.

The three-point correlation function can be written as follows
\[
\left< \Phi _{j_1} (x_1 |z_1) \Phi _{j_2}  (x_2 |z_2)  \Phi _{j_3}  (x_3 |z_3) \right> 
= {\cal A} _{j_1,j_2,j_3} 
\]
where
\begin{eqnarray}
{\cal A} _{j_{1},j_{2},j_{3}}=\prod_{r<s}\left| z_{r}-z_{s}\right|
^{2(\sum _{t=1} ^3 \Delta _{j_t} -2 \Delta _{j_r} -2 \Delta _{j_s} )}
\left| x_{r}-x_{s}\right| ^{2(2j_r +2j_s-\sum _{t=1} ^3 j_t )} C(j_{1},j_{2},j_{3})  
\label{cacarulo2}
\end{eqnarray}
with $r,s,t \in \{1,2,3 \}$ and
\[
C(j_{1},j_{2},j_{3})= \frac {1}{2\pi ^3 b^2} 
\left( \lambda ^{-1}  \pi \frac{\Gamma \left(1-b^2 \right) }
{\Gamma \left( 1+b^2 \right) }\right) ^{2+\sum _ rj_r} 
\frac{G (1+\sum _{s=1} ^3 j_{s})}{G (-1)} \prod _{r=1}^{3}   
\frac {G (-2j_r+\sum _{s=1} ^3 j_{s})}{G (1+2j_{r})}
\]
In the expression above, $b^{-2}=(k-2)$, while $\lambda $ is the coupling constant 
of the screening charge in $SL(2,\RR )_k$ WZNW model, which basically corresponds 
to the string coupling $g_s^{-2}$ \cite{note}. 
The $G (x)$ functions are defined as follows
\begin{equation}
G (x) = b^{-b^{2}x^{2}-(1+b^{2})x} \Upsilon  ^{-1}(-bx)  \label{ide}
\end{equation}
where
\[
\log \Upsilon  (x)=\frac{1}{4}\int_{0}^{\infty }\frac{d\tau }{\tau }\left(
b+b^{-1}-2x\right) ^{2}e^{-\tau }-\int_{0}^{\infty }\frac{d\tau }{\tau }%
\frac{\sinh ^{2}\left( \frac{\tau }{4}(b+b^{-1}-2x)\right) }{\sinh \left( 
\frac{b\tau }{2}\right) \sinh \left( \frac{b^{-1}\tau }{2}\right) } 
\]
The $\Upsilon (x)$ functions were introduced in reference \cite{zz} 
in the context of Liouville theory, and we will make use of them below. 
These special functions have zeroes in the lattice 
\begin{eqnarray}
x \in -b\ZZ_{\geq 0}-b^{-1}\ZZ_{\geq 0} \ , \ \ x \in b\ZZ_{>0}+b^{-1}\ZZ_{>0}  \label{lcdtm}
\end{eqnarray}
and satisfy the remarkable functional relation
\begin{equation}
\Upsilon (x+b^{\pm 1}) = \gamma (b^{\pm 1} x ) b^{\pm 1\mp 2b^{\pm} x} \Upsilon (x)  \label{shift}
\end{equation}
where $\gamma (x) = \Gamma (x) / \Gamma (1-x)$.

%...Luego, la nota de Gaetano:

%**********************************************
%%%%%%%%%%%%%%%%%%%%%%%%%%%%%%%%%%%%%%%%%%%%%

%\noindent
%Equivalently
%$$
%C(j_1,j_2,j_3) =
%\frac{1}{2 \pi^3 b^2} \frac{1}{G(-1)} 
%\left(
%\frac{\gamma(b^2)}{\pi} b^{2b^2} 
%\right)^{- \sum_{s=1}^3 j_s - Q/b}
%\pi^{1-1/b^2} b^{2/b^2} \gamma(b^2)^{-1+1/b^2} 
%$$
%\begin{equation}
%\times
%\frac{1}{ \gamma( - \sum_{s=1}^3 j_s - Q/b ) 
%\, \Upsilon( - b \sum_{s=1}^3 j_s - Q ) }
%\prod_{r=1}^3 
%\frac{ \Upsilon(-2 b j_r) }
%{\gamma(- b^2 ( 2 j_r + 1) )
%\Upsilon(-b (\sum_{s=1}^3 j_s - 2 j_r ) )}
%\end{equation}

%%%%%%%%%%%%%%%%%%%%%%%%%%%%%%%%%%%%%%%%%%%%%%%%%%%%%%%%
\section{Logarithmic fields and the operator-valued relations}

%****In this section, we will prove the operator-valued relations.

\noindent
To begin with, let us introduce the special operators 
${\Phi '} _ j (x|z)$ defined as
\[
 {\Phi '} _ j (x|z) = \frac 12 \frac{\partial}{\partial j} {\Phi } _ j (x|z)\,.
\]
It is easy to observe that these operators together with $\Phi_j$ 
form a Jordan block with respect to the Virasoro algebra, 
which is encoded in the following operator product expansions \cite{GastonJ}
\[
T(z)  {\Phi } _ j (x|w) = \frac {\Delta _j }{(z-w)^2}  {\Phi } _ j (x|w) + \frac {1}{(z-w)} 
\partial _w {\Phi } _ j (x|w) + ...
\]
\[
T(z) {\Phi '} _ j (x|w) = \frac {\Delta _j }{(z-w)^2} 
{\Phi '} _ j (x|w) - \frac {\delta _j }{(z-w)^2} {\Phi } _ j (x|w) 
+ \frac {1}{(z-w)} \partial _w {\Phi '} _ j (x|w) + ...
\]
where $2 \delta _j = \frac {2j+1}{k-2}$. 
These operators ${\Phi '} _ j$ are logarithmic fields in the CFT.

Consider the following discrete set of logarithmic fields
associated to the degenerate Kac-Moody highest weight operators
$\Phi_{j_{m,n}}$ 
$$
\Phi'_{j_{m,n}}(x| z) = \frac 12
\frac{\partial}{\partial j} \Phi_{j}(x| z)
|_{j=j_{m,n}}\,.
$$
These are the central elements in our discussion.
%In order to introduce the general statement we want to prove here, 
%let us first observe certain properties of the operators $\Phi _j $ and $\Phi ' _j $ 
%which will lead us to show the operator-valued relation we are interested on: 
The first crucial fact is that 
$$
\bar K^{\pm}_{m,n} K^{\pm}_{m,n} \Phi'_{j^{\pm}_{m,n}}(x,\bar x| z, \bar z) 
$$
is a Kac-Moody highest weight operator.
 
The proof of this statement is similar to the one presented in \cite{Zamo}
in Liouville field theory.
Let us consider $\bar K^+_{m,n} \Phi_{j}$. In a 
neighbourhood of $j = j^+_{m,n}$,
$\bar K^+_{m,n} \Phi_{j} = (j - j^+_{m,n}) A_{m,n} +
{\cal O}((j-j^+_{m,n})^2)$, where $A_{m,n}$ is an operator
of dimension $( \Delta_{m,n} + m(n-1),\Delta_{m,n} \,)$ 
and charge $j^+_{m,n}-m$ under $\bar J^3_{0}$
\cite{KacKazhdan,MalikovFF}.
$A_{m,n}$ is not a left Kac-Moody primary operator
any more but it is still a right Kac-Moody highest 
weight operator.
It follows that $K^+_{m,n} A_{m,n} = 
K^+_{m,n} \bar K^+_{m,n} \Phi'_{j^+_{m,n}}$
is also a right Kac-Moody highest weight operator.
This is because $A_{m,n}$ has the same charge under
$J^3_0$ and right dimension as 
$\Phi_{j^+_{m,n}}$.
Inverting the roles of $K^+_{m,n}$ and $\bar K^+_{m,n}$,
one can conclude that $K^+_{m,n} \bar K^+_{m,n} \Phi'_{j^+_{m,n}}$
is also a left Kac-Moody highest weight operator,
of dimension  $( \Delta_{m,n} + m(n-1), \Delta_{m,n} + m(n-1)\,)$ 
and charge $( j^+_{m,n}-m, j^+_{m,n}-m \,)$ 
under $\bar J^3_{0}, J^3_0$.
%{\bf What is a possible candidate for this state?}

The second fact is that these are precisely 
the dimension and charge of $\Phi_{\tilde \jmath^+_{m,n}}$, 
where
$$
\tilde \jmath^+_{m,n} = j^+_{m,n}-m
=  - \frac{m+1}{2} + \frac{n-1}{2}(k-2)\,.
$$
The same argument can be repeated for $j^-_{m,n}$.
In this case, the dimension and charge of
$ K^{-}_{m,n} K^{-}_{m,n} 
\Phi'_{j^-_{m,n}}(x,\bar x| z, \bar z)$ 
match those of $\Phi_{\tilde \jmath^-_{m,n}}$. 

In the following section, we will establish the operator-valued
relation
\begin{equation}
\bar K^{\pm}_{m,n} K^{\pm}_{m,n} 
\Phi'_{j^\pm_{m,n}}(x,\bar x| z, \bar z) 
= B^\pm_{m,n} \Phi_{\tilde \jmath^{\pm}_{m,n}}(x,\bar x| z, \bar z)\,, 
\label{higher1}\end{equation}
and evaluate the {\it Zamolodchikov coefficients} $B^\pm_{m,n}$.

We should mention that in the case of rational level $k-2 =p/q$ there exist two different null states in the module of $j^{\pm}_{m,n}$. This is due to the fact that $j^{\pm}_{m,n}=j^{\pm}_{m+p,n-q}$. In this particular case, we find two different operator-valued equations, one of these involving the spin $\tilde {\jmath}^{\pm}_{m+p,n-q}=\tilde {\jmath}^{\pm}_{m,n}-m$ primary field.

%6666666666666666666666666666
 
Before we proceed, let us show that the above identity 
holds in the classical limit for the operators 
$\Phi_{j^{+}_{m,1}}$ belonging to the {\it classical branch}.
This amounts to showing that 
$\partial_x^m \partial_{\bar x}^m \Phi'_{j^+_{m,1}}$
is proportional to $\Phi_{\tilde \jmath^+_{m,1}}$,
where, by (\ref{cipollina}) 
\begin{equation}
\Phi_{j^+_{m,1}} = 
\frac{m}{\pi} \left( |\gamma -x|^2 e^{\phi } + e^{-\phi } \right) ^{m-1}
\,,
\quad
\Phi_{\tilde \jmath^+_{m,1}} = 
- \frac{m}{\pi} \left( |\gamma -x|^2 e^{\phi } + e^{-\phi } \right) ^{-(m+1)}\,.  \label{esta}
\end{equation}
%First of all, we consider the following equations as 
%preliminary facts leading to the main property we want 
%to show (see (\ref{concha}) below): We have
Note that
\begin{equation}
\partial_x^m \partial_{\bar x}^m \Phi'_{j^+_{m,1}} 
= \partial_x^m \partial_{\bar x}^m 
\left(
\frac{m}{\pi}\, e^{-(m-1) \phi} A^{m-1} \log A
\right)\,,  \label{esta2}
\end{equation}
where
\begin{equation}
A(x,\bar x) = \left| \gamma -x \right| ^2 e^{2\phi } +1 \,.
\end{equation}
Then, observing that 
\begin{eqnarray}
\partial ^r _x \left( A^{r-1} \log A  \right) = 
(r-1)! A^{-1} \left( \partial _x A \right)^r \,,
%\\
%\partial _{\bar x} ^s \left( \partial _x  A \right) ^r = \frac {r!}{(r-s)!} 
%\left( \partial _x A \right)^{r-s} \left( \partial _x \partial _{\bar x}A \right)^{s}
\end{eqnarray}
and using the identities
\begin{eqnarray}
\partial _{\bar x} ^s A^{-1}  = (-1)^s s!\, A^{-s-1} 
\left( \partial _{\bar x} A \right)^{s} \,,
\quad 
\partial _{\bar x} ^s \left( \partial _x  A \right) ^r = \frac {r!}{(r-s)!} 
\left( \partial _x A \right)^{r-s} \left( \partial _x \partial _{\bar x}A \right)^{s} 
\,,
\end{eqnarray}
one finds 
\begin{equation}
\partial ^r _{\bar x} \partial ^r _{ x} \left( A^{r-1} \log A \right) =
r!(r-1)! A^{-r-1} e^{2r\phi}
\,,
\end{equation}
which implies 
%Then, by recalling that $\tilde {j} ^+ _{m,1} = -1-j^+ _{m,1}$, we eventually obtain
\begin{equation}
\partial ^m _{\bar x} \partial ^m _{ x}  \Phi ' _{j^+ _{m,1}} (x,\bar x |z,\bar z )  = 
-  m! (m-1)! \Phi _{\tilde{j}^+ _{m,1}} (x, \bar x |z, \bar z ) \,.
\label{concha}
\end{equation}
In the next section, we will evaluate the general expression of the
{\it Zamolodchikov coefficients} $B^{\pm}_{m,n}$ and we will show that
$B^+_{m,1} \rightarrow -m!(m-1)!$ in the classical limit 
$k \to \infty$.

%In summary, we proved that ...

%In fact, equation (\ref{concha}) is the classical version (a particular case) 
%of an infinite serie of operator-valued relations we will prove here. 
%These are the Kac-Moody couterpart of analogous relations recently proven by 
%Zamolodchikov for Virasoro degenerate representations in Liouville CFT. 
%The main goal of the next section is to find the generalization of 
%the factor $B^+_{m,1}=-m!(m-1)!$ in the r.h.s. of (\ref{concha}), 
%namely $B^{\pm}_{m,n}$ which we will call {\it Zamolodchikov coefficients}.

%%%%%%%%%%%%%%%%%%%%%%%%%%%%%%%%%%%%%%%%%%%%%
%%%%%%%%%%%%%%%%%%%%%%%%%%%%%%%%%%%%%%%%%%%%%
\section{Zamolodchikov coefficients and fusion rules}

Eq.(\ref{higher1}) is an operator-valued relation,
namely for every correlation function 
$$
\langle \bar K^{+}_{m,n} K^{+}_{m,n} 
\Phi'_{j^+_{m,n}}(x|z) \prod _{i=1}^{N-1} \Phi_{j_i} (x_i|z_i) \rangle 
=
B^+_{m,n} \langle  
\Phi_{\tilde \jmath^+_{m,n}}(x|z) \prod _{i=1}^{N-1} \Phi_{j_i} (x_i|z_i) 
\rangle\,.
$$
Thanks to the conformal invariance of the theory, 
it is sufficient to verify that the above equality
holds for three-point functions. 
Therefore, we will compute the quotient between
$$
{\cal A}'_{m,n} = \langle \bar K^{+}_{m,n} K^{+}_{m,n} 
\Phi'_{j^+_{m,n}} (x| z) 
\Phi_{j_1} (x_1| z_1) 
\Phi_{j_2} (x_2| z_2) 
\rangle
$$
and 
$$
\tilde {{\cal A}}_{m,n} = \langle 
\Phi_{\tilde \jmath^+_{m,n}} (x| z) 
\Phi_{j_1} (x_1| z_1) 
\Phi_{j_2} (x_2| z_2) 
\rangle\,,
$$
which will yield the explicit form of coefficient $B^+_{m,n}$.

\noindent
The latter correlator is simply given by
$$
\tilde {{\cal A}}_{m,n} = C (\, \tilde \jmath^+_{m,n},j_1,j_2 \,)
\frac{1}{|z-z_{1}|^{2(\tilde \Delta_{m,n}+\Delta_1-\Delta_2)}
|z_{12}|^{2(\Delta_1+\Delta_2-\tilde \Delta_{m,n})}
|z-z_2|^{2(\Delta_2+\tilde \Delta_{m,n}-\Delta_1)} }
$$
$$
\times \frac{1}{|x-x_{1}|^{2(\tilde \jmath^+_{m,n}+j_1-j_2+1)}
|x_{12}|^{2(j_1+j_2-\tilde \jmath^+_{m,n}+1)}
|x-x_{3}|^{2(j_2+\tilde \jmath^+_{m,n}-j_1+1)}
}\,.
$$

%Unlike the computation performed in \cite{Zamo}, the case we consider 
%here present two additional puzzles: the first is the 
%dependence on the $SL(2,\RR )$-isospin coordinates $(x,\bar x )$ and, secondly, 
%the presence of additional factors appearing (see second line of eq. (\ref{tucu}) below) 
%which are not present in 
%the Liouville case. We will see below how the consideration of the fusion rules for $\hat {sl(2)} _k$ 
%algebra enable us to handle these additional difficulties.

\noindent
Now, let us turn our attention to the three-point function ${\cal A}' _{m,n}$. 
First of all, note that $C(\, j,j_1,j_2 )$ has a first order 
zero as $j \to j^+_{m,n}$. This implies that 
$$
\langle  
\Phi'_{j^+_{m,n}} (x| z) 
\Phi_{j_1} (x_1| z_1) 
\Phi_{j_2} (x_2| z_2) 
\rangle 
$$
$$
= \frac 12 \frac{\partial C (\,j ,j_1,j_2 \,) / \partial j |_{j=j^+_{m,n}}}
{|z-z_{1}|^{2(\Delta_{m,n}+\Delta_1-\Delta_2)}
|z_{12}|^{2(\Delta_1+\Delta_2-\Delta_{m,n})}
|z-z_2|^{2(\Delta_2+\Delta_{m,n}-\Delta_1)} }
$$
$$
\times \frac{1}{|x-x_{1}|^{2(j^+_{m,n}+j_1-j_2+1)}
|x_{12}|^{2(j_1+j_2-j^+_{m,n}+1)}
|x-x_{3}|^{2(j_2+j^+_{m,n}-j_1+1)}
}\,.
$$
%%%%%%%%%%%%%%%%%%%
The operators $K^{+}_{m,n}$ and $\bar K^{+}_{m,n}$
change both the $z$ and $x$ dependence of
the correlation function. They act on the 
holomorphic and antiholomorphic factors separately.
%In particular, $K^{+}_{m,n}$ gives rise to 
%a certain differential operator 
%$$
%{\cal K}^{+}_{m,n} = ....
%$$
Since the action of $K^{+}_{m,n}$ on a right Kac-Moody
highest weight operator of charge $j^+_{m,n}$ 
and conformal dimension $\Delta_{m,n}$ 
produces another right Kac-Moody
highest weight operator of charge $\tilde \jmath^+_{m,n} =
j^+_{m,n}-m$ and conformal dimension $\tilde \Delta_{m,n} 
= \Delta_{m,n}+ m(n-1)$, we have that
$$
{K}^{+}_{m,n} \left[
(z-z_{1})^{-(\Delta_{m,n}+\Delta_1-\Delta_2)}
(z_{12})^{-(\Delta_1+\Delta_2-\Delta_{m,n})}
(z-z_2)^{-(\Delta_2+\Delta_{m,n}-\Delta_1)} 
\right.
$$
$$
\left.
\times (x-x_{1})^{-(j^+_{m,n}+j_1-j_2+1)}
(x_{12})^{-(j_1+j_2-j^+_{m,n}+1)}
(x-x_{3})^{-(j_2+j^+_{m,n}-j_1+1)}
\right]
$$
$$
= P_{m,n}(j^+_{m,n},j_1,j_2) \,\left[
(z-z_{1})^{-(\tilde \Delta_{m,n}+\Delta_1-\Delta_2)}
(z_{12})^{-(\Delta_1+\Delta_2-\tilde \Delta_{m,n})}
(z-z_2)^{-(\Delta_2+ \tilde \Delta_{m,n}-\Delta_1)} 
\right.
$$
$$
\left.
\times (x-x_{1})^{-(\tilde \jmath^+_{m,n}+j_1-j_2+1)}
(x_{12})^{-(j_1+j_2-\tilde \jmath^+_{m,n}+1)}
(x-x_{3})^{-(j_2+\tilde \jmath^+_{m,n}-j_1+1)}
\right]\,,
$$
where the function $P_{m,n}(j^+_{m,n},j_1,j_2)$ 
is given by \cite{AwataYamada}
$$
P_{m,n}(j^+_{m,n},j_1,j_2)
= \prod_{r=0}^{m-1} \prod_{s=0}^{n-1}
\left( j^+_{m,n} + j_1 - j_2 - r - s (k-2) \right)
$$
\begin{equation}
\times
\prod_{r=1}^{m} \prod_{s=1}^{n-1}
\left( - j^+_{m,n} + j_1 + j_2 + r + s (k-2) \right)\,,
\label{Pmn}\end{equation}
and the equation $P_{m,n}(j^+_{m,n},j_1,j_2)=0$
yields precisely the fusion rules for the degenerate 
Kac-Moody primary $\Phi^+_{m,n}$ \cite{AwataYamada}.
Repeating this argument for $\bar { K}^{+}_{m,n}$,
we find that 
$$
{\cal A}'_{m,n} 
= \frac{P^2_{m,n}(j^+_{m,n},j_1,j_2)\, \frac 12 \partial C (\,j ,j_1,j_2 \,) / 
\partial j |_{j=j^+_{m,n}}}
{|z-z_{1}|^{2(\tilde \Delta_{m,n}+\Delta_1-\Delta_2)}
|z_{12}|^{2(\Delta_1+\Delta_2- \tilde \Delta_{m,n})}
|z-z_2|^{2(\Delta_2+ \tilde \Delta_{m,n}-\Delta_1)} }
$$
$$
\times \frac{1}{|x-x_{1}|^{2(\tilde \jmath^+_{m,n}+j_1-j_2+1)}
|x_{12}|^{2(j_1+j_2- \tilde \jmath^+_{m,n}+1)}
|x-x_{3}|^{2(j_2+ \tilde \jmath^+_{m,n}-j_1+1)}
}\,.
$$

%%%%%%%%%%%%%%%
%\subsection{}%
%%%%%%%%%%%%%%%

\noindent
Then we write the quotient of both correlators as follows
$$
\frac{{\cal A}'_{m,n}}{\tilde {{\cal A}}_{m,n} P^{\,2}_{m,n}(j^+_{m,n},j_1,j_2) } 
= \frac 12 \frac{ \partial C (\,j ,j_1,j_2 \,) / \partial j |_{j=j^+_{m,n}}}
{C (\,\tilde \jmath^+_{m,n} ,j_1,j_2 \,)}
= 
-b \left(
\frac{\lambda \gamma(b^2)}{\pi} b^{2b^2} 
\right)^{( \tilde \jmath_{m,n} - j_{m,n} )} 
\frac{\Upsilon'(- 2 b j_{m,n})\,}{\Upsilon(- 2 b \tilde \jmath_{m,n})\,}
$$
$$
\times
\frac{\gamma( - (\tilde \jmath_{m,n} + j_1 + j_2 + k-1 ))\,
\gamma( -b^2 ( 2 \tilde \jmath_{m,n} + 1 )  )\,}
{\gamma( - (j_{m,n} + j_1 + j_2 + k-1))\,
\gamma( -b^2 ( 2 j_{m,n} +1 ) )\,} 
$$
\begin{equation}
\times
\frac{
\Upsilon(-b(\tilde \jmath_{m,n} + j_1 + j_2 + k-1) )\, 
\Upsilon(-b(\tilde \jmath_{m,n} + j_1 - j_2) )\,
\Upsilon(-b(j_1 + j_2 - \tilde \jmath_{m,n}) )\,
\Upsilon(-b(j_2 + \tilde \jmath_{m,n} - j_1) )\,
}
{
\Upsilon(-b( j_{m,n} + j_1 + j_2 + k-1) )\, 
\Upsilon(-b( j_{m,n} + j_1 - j_2) )\,
\Upsilon(-b(j_1 + j_2 - j_{m,n}) )\,
\Upsilon(-b(j_2 + j_{m,n} - j_1) )\,
}\,,  \label{tucu}
\end{equation}
where $Q = b + b^{-1}$ and $\Upsilon ' (x) = \frac {d \Upsilon }{dx} (x)$. 
Here we wrote the expression in terms of the $\Upsilon (x)$ functions introduced before.

\noindent
The last term in (\ref{tucu}) is equal to
$$
\frac{
\Upsilon(-b(\tilde \jmath_{m,n} + j_1 + j_2 + k-1) )\, 
\Upsilon(-b(\tilde \jmath_{m,n} - ( j_1 + j_2 + k-1 ) ) )\,
%\Upsilon(-b(\tilde \jmath_{m,n} + j_1 - j_2) )\,
%\Upsilon(-b(\tilde \jmath_{m,n} + j_2 - j_1) )\,
}
{
\Upsilon(-b( j_{m,n} + j_1 + j_2 + k-1) )\, 
\Upsilon(-b( j_{m,n} - ( j_1 + j_2 + k-1 ) ) )\,
%\Upsilon(-b( j_{m,n} + j_1 - j_2) )\,
%\Upsilon(-b( j_{m,n} +j_2 - j_1) )\,
}\,
$$
\begin{equation}
\times
\frac{
\Upsilon(-b(\tilde \jmath_{m,n} + j_1 - j_2) )\,
%\Upsilon(-b(j_1 + j_2 - \tilde \jmath_{m,n}) )\,
\Upsilon(-b(\tilde \jmath_{m,n} + j_2 - j_1) )\,
}
{
\Upsilon(-b( j_{m,n} + j_1 - j_2) )\,
%\Upsilon(-b(j_1 + j_2 - \jmath_{m,n}) )\,
\Upsilon(-b( j_{m,n} +j_2 - j_1) )\,,
}
\label{cippolone}\end{equation}
where we used the identity
$\Upsilon(Q-x) = \Upsilon(x)$.

By using (\ref{shift}), it can be also proven that
$$
\frac{
\Upsilon(-b ( \tilde \jmath_{m,n} + x ) )\, 
\Upsilon(-b ( \tilde \jmath_{m,n} - x ) )\,
}
{
\Upsilon(-b ( j_{m,n} + x ) )\,
\Upsilon(-b ( j_{m,n} - x ) )\,
} = \frac{(-1)^{nm}}{p^2_{m,n}(x)} \,,
$$
where
\begin{equation}
p_{m,n}(x) =
b^{mn} \prod_{r,s} \left( x - \frac{r}{2} - \frac{s}{2}\, (k-2) \right)\,,
\label{identity1}\end{equation}
and $r = \{ -m+1, -m+3 \ldots, m-3, m-1 \}$,
$s = \{-n+1, -n+3 \ldots, n-3, n-1\}$.
Then, by (\ref{identity1}), 
equation (\ref{cippolone}) becomes 
$$
\frac{1}{ b^{4 m n } \,\tilde P^2( j^+_{m,n},j_1,j_2 ) }\,, 
$$
where
$$
\tilde P_{m,n}(j^+_{m,n},j_1,j_2)
= \prod_{r=0}^{m-1} \prod_{s=0}^{n-1}
\left( j^+_{m,n} + j_1 - j_2 - r - s (k-2) \right)
$$
\begin{equation}
\times
\prod_{r=1}^{m} \prod_{s=1}^{n}
\left( - j^+_{m,n} + j_1 + j_2 + r + s (k-2) \right)\,.
\label{tildePmn}\end{equation}
%%%%%%%%%%%%%%%%%%%%%
\noindent
On the other hand, by using standard formulae involving $\Gamma (x)$ functions, we also find
$$
R_{m,n} \equiv
\frac{\gamma( - (\tilde \jmath_{m,n} + j_1 + j_2 + k-1 ))\,}
{\gamma( - (j_{m,n} + j_1 + j_2 + k-1))\,} 
=
\frac{\gamma( - (j_{m,n} + j_1 + j_2 + k-1 ) + m )\,}
{\gamma( - (j_{m,n} + j_1 + j_2 + k-1))\,} 
$$
$$
%= (-1)^m \left( \prod_{i=0}^{m-1} 
%(- (j_{m,n} + j_1 + j_2 + k-1 ) + i ) 
%\right)^2 
=
(-1)^m \left( \prod_{i=0}^{m-1} 
( j_{m,n} + j_1 + j_2 + k-1 - i ) 
\right)^2\,. 
$$

%%%%%%%%%%%%%%%%%%%%%%
\noindent
Finally,
$$
\frac{{\cal A}'_{m,n}}{\tilde {{\cal A}}_{m,n} P^{2}_{m,n}(j^+_{m,n},j_1,j_2) } 
= \frac 12   \frac{ \partial C (\,j ,j_1,j_2 \,) / \partial j |_{j=j^+_{m,n}}}
{C (\,\tilde \jmath^+_{m,n} ,j_1,j_2 \,)}
= - 
\left(
\frac{\lambda \gamma(b^2)}{\pi} b^{2b^2} 
\right)^{-m} 
\frac{\Upsilon'(- 2 b j_{m,n})\,}{\Upsilon(- 2 b \tilde \jmath_{m,n})\,}
$$
$$
\times
\frac{\gamma( -b^2 ( 2 \tilde \jmath_{m,n} + 1 )  )\,}
{\gamma( -b^2 ( 2 j_{m,n} +1 ) )\,} 
\, \frac{R_{m,n}}{b^{4mn-2} \tilde  P^2_{m,n}(j^+_{m,n},j_1,j_2) }
$$
\begin{equation}
= \left(
\frac{\lambda \gamma(b^2)}{\pi} b^{2b^2} 
\right)^{-m} 
\frac{\Upsilon'(- 2 b j_{m,n})\,}{\Upsilon(- 2 b \tilde \jmath_{m,n})}
\frac{\gamma( -b^2 ( 2 \tilde \jmath_{m,n} + 1 )  )\,}
{\gamma( -b^2 ( 2 j_{m,n} +1 ) )\,} 
\, \frac{b^{2-4mn} (-1)^{m+1} }{P^2_{m,n}(j^+_{m,n},j_1,j_2) }\,,
\end{equation}
where we used
$$
\frac{ \tilde P^2_{m,n} }{ P^2_{m,n} } = (-1)^m R_{m,n}\,.
$$

\noindent
Therefore, we find that any dependence 
of the ratio ${\cal A}'_{m,n} / \,\tilde {{\cal A}}_{m,n}$
on $j_1,j_2$ drops out and we obtain
\begin{equation}
B^+_{m,n} = \frac{{\cal A}'_{m,n}}{\tilde {{\cal A}}_{m,n}}
=
\left(
\frac{\lambda \gamma(b^2)}{\pi} b^{2b^2} 
\right)^{-m} 
\frac{\Upsilon'(- 2 b j^+_{m,n})\,}{\Upsilon(- 2 b \tilde \jmath^+_{m,n})}
\frac{\gamma( -b^2 ( 2 \tilde \jmath^+_{m,n} + 1 )  )\,}
{\gamma( -b^2 ( 2 j^+_{m,n} +1 ) )\,} 
\,
\times b^{2-4mn} (-1)^{m+1} \,.
\label{temp1B+mn}\end{equation} 
The calculation for the case $j^-_{m,n}$ follows the same lines as 
the previous one. Then, we find
\begin{equation}
B^-_{m,n} 
= 
\left(
\frac{\lambda \gamma(b^2)}{\pi} b^{2b^2} 
\right)^{m} 
\frac{\Upsilon'(- 2 b j^-_{m,n})\,}{\Upsilon(- 2 b \tilde \jmath^-_{m,n})}
\frac{\gamma( -b^2 ( 2 \tilde \jmath^-_{m,n} + 1 )  )\,}
{\gamma( -b^2 ( 2 j^-_{m,n} +1 ) )\,} 
\,
\times b^{2-4m(n-1)} (-1)^{m+1} \,.
\label{temp1B-mn}\end{equation} 
In order to compare the above equations with the analogous
results obtained in \cite{Zamo} for Liouville CFT, 
one can further simplify the above expressions and write
\begin{equation}
B^+_{m,n} = \left(
\frac{ \lambda \gamma(b^2)}{\pi} \right)^{-m} (-1)^{m+1} b^{2-2n-4mn} 
\gamma (n+mb^2) \prod _{p=1-n} ^{p=n-1} \prod _{q=1-m} ^{q=m-1} (pb^{-1}+qb)\,,   
\label{saxo}
\end{equation}
where $(p,q) \ne (0,0)$.

\noindent
Note that the classical limit, $b\rightarrow 0$,
for the Zamolodchikov coefficient in {\it classical branch} is
\begin{equation}
\lim _{b\rightarrow 0} \, B^+ _{m,1} = - (\lambda ^{-1} \pi )^m m! (m-1)!  \label{v67}
\end{equation}
which, upon fixing $\lambda = \pi $, exactly agrees with the classical result 
obtained in (\ref{concha}). Thus we see that the exact result (\ref{saxo}) 
is fully consistent with the classical limit. 
Besides, even at finite $b$ we have $B^+_{1,1}=-\frac {\pi }{\lambda }$. 

Finally, let us notice that Eq. (\ref{concha}), which is the classical limit of (\ref{higher1}) in the case $(m,n)=(1,1)$, can be rewritten as the Liouville equation $\partial _x \partial _{\bar x} \varphi (x|z) = e^ {-2\varphi (x|z)}$ for the field $\varphi (x|z) = \log \Phi _{j_{2,1}}(x|z)$ in terms of the {\it boundary} variables $(x,\bar x)$.

%\newpage

%%%%%%%%%%%%%%%%%%%%%%%%%%%%%%%%%
\section{Conclusions}

In this letter, we derived an infinite set of operator-valued relations, Eq. (\ref{higher1}), which hold for degenerate representations of $\hat {sl(2)} _k$ Kac-Moody algebra.
These relations are similar to those recently found by Zamolodchikov for Virasoro degenerate representations in Liouville conformal field theory \cite{Zamo}. 

By studying the functional form of the three-point functions on the sphere and considering the fusion rules of $\hat {sl(2)}_k$ algebra we were able to find the explicit expression of the Zamolodchikov coefficients (\ref{temp1B+mn})-(\ref{temp1B-mn}) for this non-rational CFT.

The operator-valued relations translate into differential equations satisfied by correlation functions involving particular Kac-Moody primary states. These are equations in terms of the $SL(2,\RR )$-isospin variables $(x,\bar x)$. Furthermore, the first equation of this infinite set resembles the Liouville equation. This observation could be relevant in the context of the $AdS_3/CFT_2$ correspondence since the variables $(x,\bar x)$ precisely represent the coordinates of the boundary, where the dual conformal field theory is formulated. This could be an interesting topic for further research.

%Besides, we discussed the large $k$ limit of the obtained formula and we verified that this leads to recover the expected behaviour (\ref{v67}), consistent with the properties of wave functions in the classical regime.

%%%%%%%%%%%%%%%%%%%%%%%%%%

%%%%%%%%%%%%%%%%%%%%%%%%%%%
\newpage

%%%%%%%%%%%%%%%%%%%%%%%%%%%
\section*{Acknowledgements}
We would like to thank Stefano Bolognesi, Juan Maldacena, Marco Matone, Luca Mazzucato, Yu Nakayama, J\"org Teschner and David Shih for discussions and comments.
G. Bertoldi is supported by the Foundation BLANCEFLOR
Boncompagni-Ludovisi, n\'ee Bildt.
G. Giribet is supported by Institute for Advanced Study and Fundaci\'on Antorchas; on leave from University of Buenos Aires.

%%%%%%%%%%%%%%%%%%%%%%%%%%%

\end{document}